# A new method for determining the filled point of the tooth by Bit-Plane Algorithm

Zahra Alidousti
Department of computer engineering
University of Isfahan
Isfahan, Iran
Zahra_Alidousti@yahoo.com

Maryam Taghizadeh Dehkordi
Faculty of Technology and Engineering
Shahrekord University
Shahrekord, Iran
Maryam_td121@yahoo.com

*Abstract*—Up to now, researchers have applied segmentation techniques in their studies on teeth images, with a construction on tooth root length and depth. In this paper, a new approach to the exact identification of the filled points of tooth is proposed. In this method the filled teeth are detection by applying the Bit-Plane algorithm on the OPG images. The novelty of the proposed method is that we can use it in medicine for detection of dental filling and we calculate and present the area of the filled points which may help dentists to assess the filled point of the tooth. The experimental results, confirmed by the dentists, clearly indicate that this method is able to separate the filled points from the rest of healthy teeth completely.

Keywords—Dental filling; Image segmentation; Bit-Plane algorithm; Dental Image Processing; Medical Images.

I. INTRODUCTION

One of the most considerable challenges in the field of dental health is the detection of defected points which gradually may lead to teeth decay. Dentists can detect and treat dental cavities by viewing X-Ray images, which there exists, the problem that occurs after filling the decayed point which may need detection point in the subsequent photographs of the tooth. Typically, the filled points fail after a period of time and the cavities are revealed again which can cause pain and discomfort in the patient. At this phase, accurate detection of cavities points becomes essential.

In [1], the K-Means algorithm is applied for detection of cavities points on 1 or 2 teeth images not investigated OPG images. In [2] pixel color technique is used to detect the filled points and the gap size between the teeth on the RGB images. Bhan Anupama et al. [3] use the Top-Hat algorithm to identify the tooth cavities on X-Ray images. The advantage of this method is the identification of the cavities, which assists the dentist to identify defected points that do not have an adverse effect on the root of the tooth in a more accurate manner. The segmentation techniques are applied to test root and multi-tooth length in [4]. In [5], the focus is only on some of the points on the tooth indicate that segmentation is not accurate, and as noted in this article, this method for detecting tooth cavities is inefficient. According to [6], filled points are not precisely identified, while only filled points on adjacent teeth are identified. The run studies, that are so far have identified, are segmented the damages tooth as to detect the root depth with no detection on the whole tooth. In [7], image enhancement is run by watershed and modification of kernel function. This method is not standard in detecting the features of x-ray images. The machine learning classification technique is run by Olsen, Grace F. et al. [8] and the focus is to identify which pixels are classified using the pixel feature vectors. In an experimental result in [9], the focus is on segmentation between both the tooth and alveolar bone and it is important that tooth segmentation methods are presented in the 3D models of tooth and alveolar bones which is required for orthodontic treatment. As demonstrated in [10], the k-means algorithm is run for detecting the cavities and decay tooth. As the results show the root length and cavities are not detected completely, because experimental is through on the surface of the teeth.

According that all points on the OPG images are concern, in this study, for the first time the Bit-Plane algorithm is applied on OPG images to detect the filled point of the tooth and the statistical parameter like area is calculated.

This article is structured as follow: the proposed method is presented in sec.2; the experimental results are expressed in sec.3; and the conclusion is concluded in sec.4.

## II. PROPOSED METHOD

An algorithm is proposed for the initial segmentation of the images shot to accurately identify points of filled dentures.

For this purpose, first, the obtained images are fed in to Top-Hat algorithm in order to generate Input Original Images, and next, the obtained images are fed into Bit-Plane algorithm in order to generate a target images. By applied Bit-Plane technique the image is divided into eight pages, where changes are made on the 7$^{th}$ and 8$^{th}$ pages. The reason for making changes only on these two pages is due to the high value in the logical calculations because they have the most valuable bits in converting pixels of the image in to binary numbers; therefor shifting to valuable bits causes a clear and logical contradiction between pixels and facilitates the objective of separating specific parts of the image, Fig.1.

number between 0 and 255, each pixel can be represented by an eight bit binary number and all of the bits with different positions make planes of one to eight, Fig.2.

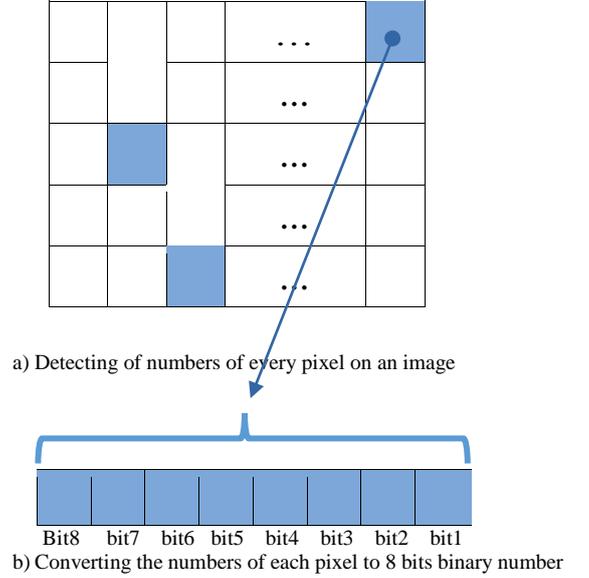

a) Detecting of numbers of every pixel on an image

b) Converting the numbers of each pixel to 8 bits binary number

Fig.2. An image is including pixels (a), every pixel convert to 8bits (b)

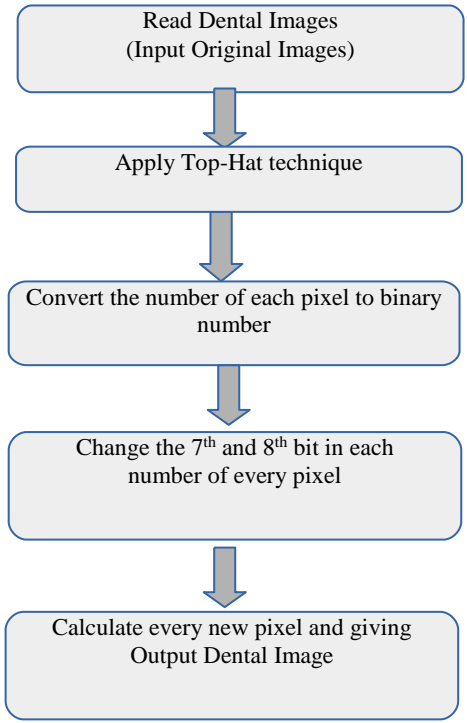

Fig.1. Bit-Plane Algorithm flowchart for image segmentation in the proposed method

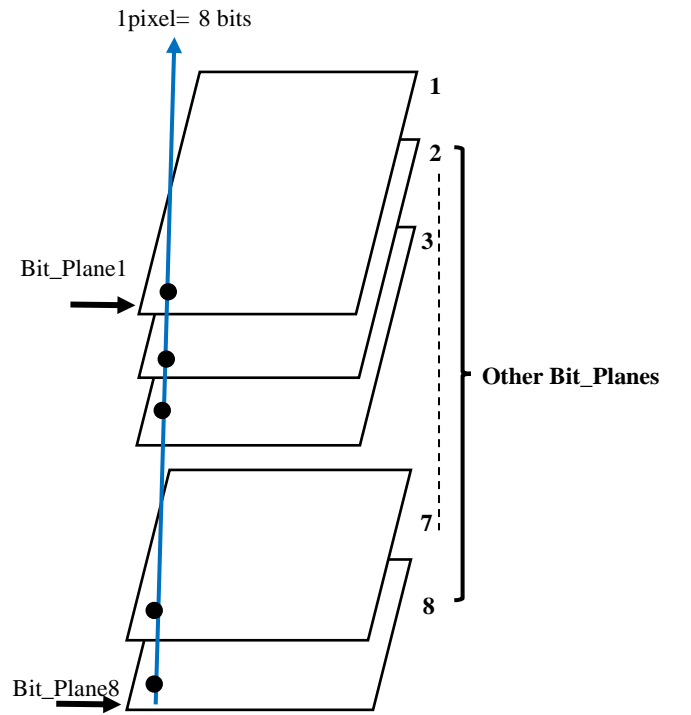

Fig.3. General Structure of the proposed algorithm

As it is illustrated in Fig.1, in the second stage the Top-Hat technique is performed. It can extract small elements and details from Input Original Images that are in previous stage.

In the third stage, it is noticeable to calculating each pixel of given images. These images are seen as a two dimensional matrix and every cell in matrix show a number that can represent pixel value. Because pixel values are decimal

As it is shown in Fig.3, changing 7th and 8th bits on 7th and 8th planes can produce a new value for every pixel and this changing is influencing for detecting some points of images that are our target to detection.

where parts of the teeth with a mild filling are completely visible through the provided method, Fig.4. Also the area of filled points of dental are extracted from the Original images and the obtained images after applying our method (in pixels). Based on the parameter the detection of the area of filled points is done more carefully, Table.1, 2.

### III. EXPERIMENTAL RESULTS

Experiments are performed and verified with MATLAB. The filled dentures on the OPG images are well illustrated

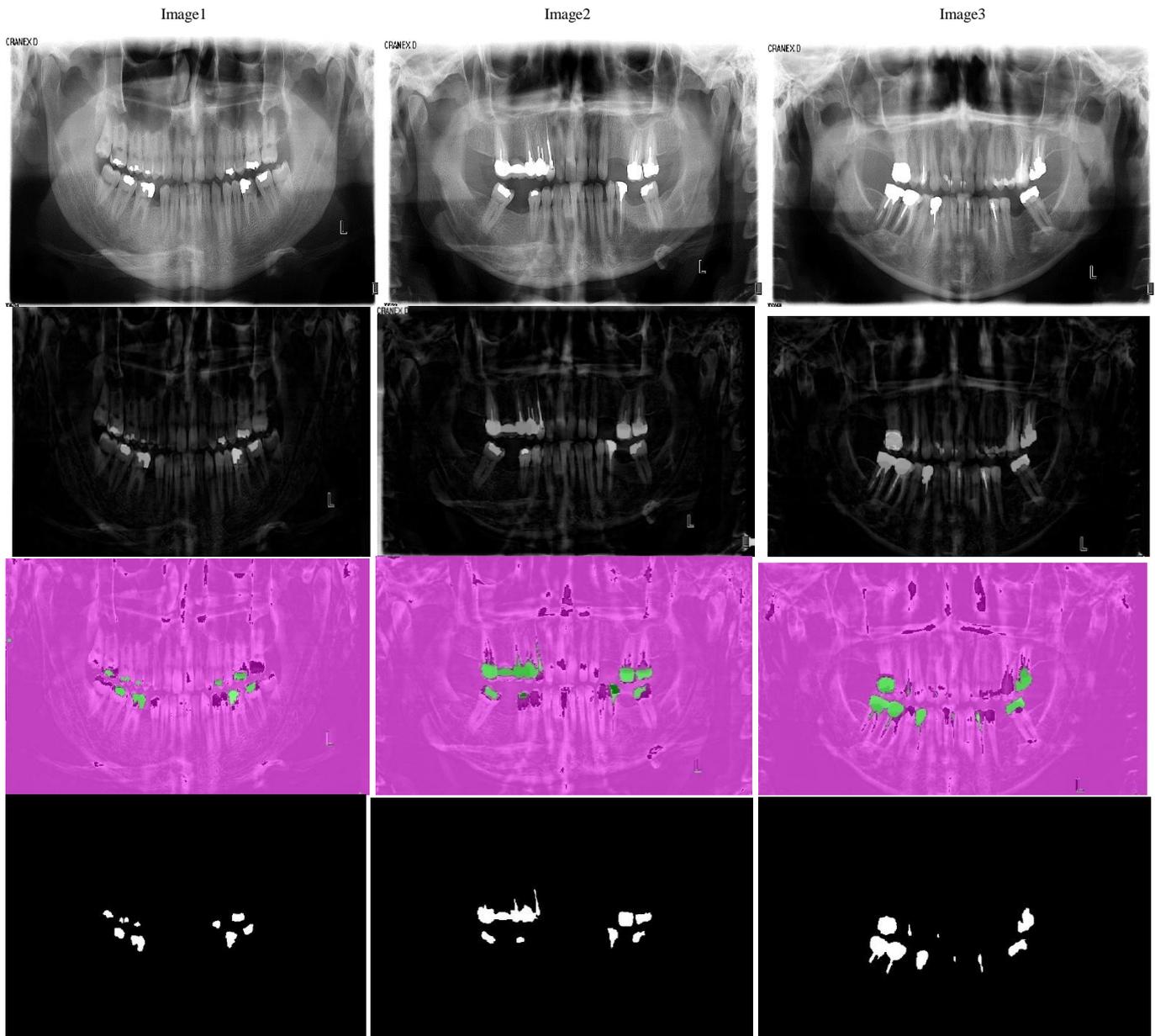

Fig.4. The results of Top_Hat technique and Bit_Plane algorithm and performing threshold technique on three samples. Original Images (first row), given images by Top_Hat technique (second row), given images by Bit-Plane algorithm (third row), given images after thresholding (forth row).

TABLE 1. Calculating of the area of filled point for image1, 2, 3

| Test Image | Area of filled point of the tooth (proposed method) (mm$^2$) | Area of filled point of the tooth (manually) (mm$^2$) | Error ( % ) (mm$^2$) |
|---|---|---|---|
| Image1 | 2.4304 | 1.96 | 0.4704 |
| Image2 | 15.9936 | 13.4064 | 2.5872 |
| Image3 | 5.5664 | 4.9392 | 0.6272 |
|  | 7.9968 | 6.7685 | 1.2282 |

TABLE 2. Comparison of the proposed method and previous methods to calculated the area of filled points of dental

| Test Image | Algorithm | Area (Mean) |
|---|---|---|
| [3] | Top_Hat and sharped filter | 4.0789 |
| [1] | K_means Algorithm | 4.1531 |
| [2] | Clustering Based Segmentation | 6.60 |
| Our Method | Bit-Plane Algorithm | 7.9968 |

Table 1 shows the area of filled points of dental images using the proposed method, the area determined manually and its error. The dimension are mm$^2$ with the assumption every pixel area is 0.0784 mm$^2$ [4].

Table 2 compares the mean of the errors using previous and our method. This results confirm that the proposed approach has been able to identify the filling points of dental images more accuracy.

CONCLUSION

As explained in the study, the objective here is to identify all filled points of a tooth represented in given images. This is accomplished by applying the Bit-Plane algorithm. This algorithm, by shifting onto valuable pages is able to segment the OPG images in an accurate manner and the area of dental filled points are visible exactly. Therefore, the dentist can accurately observe all the filled points by viewing the output images completely.

ACKNOWLEDGMENT

We would like to show our gratitude to Resalat-Specialist Dental Clinic of Shahrekord, Iran for providing teeth images to use, as well as Dr. Nader Rahimi who approved the obtained results.

REFERENCES

[1] Patil, Harshada, and Sagar A. More. "Comparative study of Root Canal Treatment and Caries Detection in Dentistry." (2017).

[2] Osadcha, Oleksandra, Agata Trzcionka, Katarzyna Pachońska, and Marek Pachoński. "Detection of Dental Filling Using Pixels Color Recognition." In *International Conference on Information and Software Technologies*, pp. 347-356. Springer, Cham, 2018.

[3] Bhan, Anupama, Garima Vyas, Sourav Mishra, and Pulkit Pandey. "Detection and Grading Severity of Caries in Dental X-ray Images." In *Micro-Electronics and Telecommunication Engineering (ICMETE), 2016 International Conference on*, pp. 375-378. IEEE, 2016.

[4] Purnama, I. Ketut Eddy, Ima Kurniastuti, Margareta Rinastiti, and Mauridhi Hery Purnomo. "Semi-automatic determination of root canal length in dental X-ray image." In *Instrumentation, Communications, Information Technology, and Biomedical Engineering (ICICI-BME), 2015 4th International Conference on*, pp. 49-53. IEEE, 2015.

[5] Datta, Soma, and Nabendu Chaki. "Detection of dental caries lesion at early stage based on image analysis technique." In *Computer Graphics, Vision and Information Security (CGVIS), 2015 IEEE International Conference on*, pp. 89-93. IEEE, 2015.

[6] Gayathri, V., Hema P. Menon, and Amrita Viswa. "Challenges in Edge Extraction of Dental X-Ray Images Using Image Processing Algorithms-A Review." (2014).

[7] W. Kuang and W.Ye, "A Kernel-Modified SVM Based Computer-Aided Diagnosis System in Initial Caries," in Intelligent Information Technology Application, 2008. IITA'08. Second International Symposium on, 2008, pp. 207-211.

[8] Olsen, Grace F., et al. "An image-processing enabled dental caries detection system." *Complex Medical Engineering, 2009. CME. ICME* International Conference on. IEEE, 2009.

[9] Gan, Yangzhou, et al. "Tooth and alveolar bone segmentation from dental computed tomography images." IEEE journal of biomedical and health informatics 22.1 (2018): 196-204.

[10] Koutsouri, Georgia D., et al. "Detection of occlusal caries based on digital image processing." *Bioinformatics and Bioengineering (BIBE), 2013 IEEE 13th International Conference on*. IEEE, 2013.